\documentclass{PoS}
\usepackage[normalem]{ulem}
\usepackage{multirow}
\usepackage{siunitx}
\usepackage{url}

\usepackage{caption}
\usepackage{subcaption}

\newcommand{\lat}{\textit{Fermi}-LAT~}

\title{Maximum likelihood spectral fitting and its application to EBL constraints}

\ShortTitle{VERITAS EBL Constraints}

\author{\speaker{Stephan O'Brien} for the VERITAS Collaboration\footnote{https://veritas.sao.arizona.edu} \footnote{for collaboration list see PoS(ICRC2019)1177}\\
        Department of Physics, McGill University, Montreal, Canada\\
        E-mail: \email{stephan.obrien@mcgill.ca}\\

        }

\abstract{
The extragalactic background light (EBL) is the second-most-intense form of cosmic background light (the first being the cosmic microwave background) and contains the redshifted optical radiation, from infra-red to ultraviolet, emitted across all epochs, making it of great cosmological interest.
While direct measurements of the EBL are hampered by foreground contamination, observations of VHE emission from distant sources can be used to obtain indirect measurements of the EBL.
In this work a maximum-likelihood fit is applied to the energy spectra of blazars observed by VERITAS, an array of ground-based imaging atmospheric Cherenkov telescopes sensitive to very-high-energy (VHE; E>100 GeV) gamma rays.
Using theoretical models of the EBL shape and intensity, the EBL normalization is treated as a free parameter, allowing for model-dependent constraints to be obtained.
Details of this maximum-likelihood analysis and preliminary model-dependent constraints on the EBL, are presented.
}

\FullConference{36th International Cosmic Ray Conference ICRC2019\\
		July 24th - August 1st, 2019\\
		Madison, WI, U.S.A.}

\begin{document}

\section{Introduction}

The extragalactic background light (EBL) is the second most intense form of cosmological background light.
Its spectral energy distribution (SED) shows a double-peaked structure with peaks in the optical and IR wavebands.
Direct measurements of the EBL intensity are challenging due to contamination from foreground zodiacal light.
Theoretical models of the intensity and redshift evolution of the EBL have been developed (see \cite{dwek2013} for a discussion of direct measurement and modeling efforts).
In this work, the models by \cite{finke10} \cite{dominguez11} \cite{franceschini08} and \cite{gilmore12} (hereinafter Fi10, D11, F08 and G12, respectively) are considered.

\par Blazars are a sub-class of active galactic nuclei (AGN) which possess a relativistic jet.
In the case of blazars, the relativistic jet is orientated closely to the line of sight, resulting in the emission being enhanced due to Doppler boosting.

At the time of writing there are 78 AGN detected\footnote{\url{http://tevcat2.uchicago.edu}} at VHE energies, with the majority being blazars.
Being extragalactic sources, blazars can be used as cosmological probes, allowing for the study of photon propagation over cosmological scales.
Indeed, their VHE photon flux will undergo an energy and redshift dependent attenuation, due to photon-photon interactions with EBL photons such that:
\begin{equation}
    \label{eqn:pp_ebl}
    \gamma_{VHE} + \gamma_{EBL} = e^{+} + e^{-}.
\end{equation}

\noindent This attenuation of the VHE photon flux modifies the observed spectrum such that:
\begin{equation}
    \label{eqn:spectra_ebl}
    \left(\frac{d\phi}{dE}\right)_{obs} = \left(\frac{d\phi}{dE}\right)_{int} e^{-\tau(E,z)},
\end{equation}
where $\left(\frac{d\phi}{dE}\right)_{obs}$ and $\left(\frac{d\phi}{dE}\right)_{int}$ are the observed and intrinsic energy spectra and $\tau(E,z)$ is the EBL opacity.
Hence by making detailed measurements of VHE spectra, constraints on the EBL opacity can be obtained.
For details on EBL constraints using HE and VHE observations of blazars, see \cite{hess_direct, biteau, magic_recent, fermi, fermipTeV, elisa}.

\par VERITAS (\textbf{V}ery \textbf{E}nergetic \textbf{R}adiation \textbf{I}maging \textbf{T}elescope \textbf{A}rray \textbf{S}ystem) is an array of four 12-m imaging atmospheric-Cherenkov telescopes, located at the Fred Lawrence Whipple Observatory (FLWO) in southern Arizona USA (31 40N, 110 57W,  1.3km a.s.l.).
It is sensitive to gamma-ray photons in the $100\mathrm{~GeV } \mathrm{-} > 30 \mathrm{~TeV}$ energy range.
Each telescope has 345 facets, with a 499 PMT camera located at the focal plane giving a field of view of $3.5^{\circ}$.
The 68\
In its current configuration VERITAS can detect a source with flux 1\
For full details of VERITAS and its performance see \cite{VERITAS_Performance}.

\par As part of VERITAS' long-term plan, VERITAS takes dedicated deep exposures on a number of TeV blazars, across a range of redshifts.
VERITAS also regularly monitors known TeV emitters for periods of enhanced emission which would allow for the collection of high-statistics spectra.

VERITAS maintains an active target of opportunity (ToO) program, which may be self-triggered by VERITAS observations or triggered by other multiwavelength (MWL) instruments such as, for example, \lat.
This allows for the detection of enhanced MWL states and the detection of new blazars at VHE energies.
Data used in this study were taken as part of deep exposure, monitoring and ToO programs.

In these proceedings model-dependent constraints on the intensity of the EBL are obtained using VERITAS observations of VHE Blazars.
In Section \ref{sec:method} the method employed to obtain constraints on the EBL intensity is discussed. The data used in this analysis are discussed in Section \ref{sec:data}. Constraints on the EBL intensity are presented in Section \ref{sec:results} and uncertainties due to the analysis method are discussed. Finally, future prospects are discussed in Section \ref{sec:discus}.

\section{Methodology}
\label{sec:method}

\par The spectral analysis is performed using a forward-folding binned-likelihood analysis developed by \cite{piron}.
This method forward-folds the spectral model with the instrument response functions, in order to obtain a prediction of the observed counts:
\begin{equation}
    \label{EQN:counts}
    s^{pred} = T_{live} \int_{\tilde{E}_{min}}^{\tilde{E}_{max}} d\tilde{E} \int_0^{\infty} dE \left(\frac{d\phi}{dE}\right)_{obs} A_{Eff}(E) \gamma(E\rightarrow\tilde{E}),
\end{equation}{}
where $T_{live}$ is the dead time corrected exposure, $\left(\frac{dN}{dE}\right)_{obs}$ is the spectral model, $A_{Eff}$ is the effective area and $\gamma(E\rightarrow\tilde{E})$ is the probability of reconstructing a photon of true energy $E$ to have a reconstructed energy $\tilde{E}$. One can fit to the intrinsic spectrum by incorporating EBL absorption in Equation \ref{EQN:counts}:
\begin{equation}
    \left(\frac{d\phi}{dE}\right)_{obs} = \left(\frac{d\phi}{dE}\right)_{int} e^{-\alpha_{EBL}\tau(E,z)},
\end{equation}{}
where $\left(\frac{d\phi}{dE}\right)_{int}$ is the intrinsic spectrum, $\tau(E,z)$ is a model of the EBL opacity and $\alpha_{EBL}$ is an opacity scale factor such that $\alpha_{EBL} = 0$ provides no EBL absorption, and $\alpha_{EBL} = 1$ provides EBL absorption as suggested by one's EBL opacity model.

\par Constraints on $\alpha_{EBL}$ are obtained by scaling $\alpha_{EBL}$ in the range [0,2] (in steps of 0.05) and fitting to the intrinsic spectra assuming $\alpha_{EBL}$.
For the intrinsic spectra four different spectral models are considered, power law (PWL); log parabola (LP); power law with exponential cutoff (PWL-EC) and log parabola with exponential cutoff (LP-EC):\\

~\begingroup
    \fontsize{8pt}{10pt}\selectfont
\begin{eqnarray}
    &PWL &= N_0 \left(\frac{E}{E_0}\right)^{\Gamma},\\
    &LP &= N_0 \left(\frac{E}{E_0}\right)^{\Gamma + \beta\log\left(\frac{E}{E_0}\right)},\\
    &PWL-EC &=N_0 \left(\frac{E}{E_0}\right)^{\Gamma} e^{-\left(\frac{E}{E_C}\right)},\\
    &LP-EC &= N_0 \left(\frac{E}{E_0}\right)^{\Gamma + \beta\log\left(\frac{E}{E_0}\right)}e^{-\left(\frac{E}{E_C}\right)}.
\end{eqnarray}
\endgroup
The profile likelihood for each spectra and each model are obtained as a function of $\alpha_{EBL}$.
The combined profile likelihood is obtained by summing the $\log\mathcal{L}$ profiles for the most probable spectral model (see Section \ref{sec:data}) for each source.

\par The maximum-likelihood estimate (MLE) of $\alpha_{EBL}$ is obtained by finding the value which optimizes the combined profile likelihood.
To compare the MLE of $\alpha_{EBL}$ to that of a no-EBL scenario, one can invoke Wilk's Theorem:
\begin{equation}
\label{eqn:wilks}
    TS = 2\log\left(\frac{\mathcal{L}(\hat{\alpha_{EBL}})}{\mathcal{L}(\alpha_{EBL} = 0)}\right) \sim \chi^2,
\end{equation}{}
where $\hat{\alpha_{EBL}}$ is the MLE of $\alpha_{EBL}$.
Equation \ref{eqn:wilks} also allows for an estimate of the improvement to the spectral fits by including EBL attenuation:
\begin{equation}
    \label{eqn:tssigma}
    \sqrt{TS} \sim \sigma,
\end{equation}{}
where $\sigma$ can be considered the ``detection significance'' of the EBL imprint.

\par Modest constraints are applied to the spectral parameters to ensure physicality, without making the result dependent on a single blazar emission or particle acceleration model. In general the fit parameters are required such that the energy spectrum is decreasing with energy, may have a concave  --- but not convex structure --- and any cutoff must be constrained by the analysis energy. One arrives at the modest constraints of $\Gamma < -1$, $\beta < 0.$ and $E_C < 100$ TeV. The constraint on the cutoff energy is conservatively placed given no energy spectra extend beyond 20 TeV.

\par Once $\hat{\alpha_{EBL}}$ is obtained from the coarse scan of $\Delta\alpha_{EBL} = 0.05$, a finer scan of $\Delta\alpha_{EBL} = 0.01$ is performed in the  $\pm 2\sigma$ region about $\hat{\alpha_{EBL}}$.
The final values of $\hat{\alpha_{EBL}}$ reported and their corresponding uncertainty are derived from the finer scan.

\section{Data Selection}
\label{sec:data}

\begin{table}[ht]
    \centering
    \resizebox{0.65\textwidth}{!}{
    \begin{tabular}{|c||c|c|c|c|c|c|}
    \hline
    Source & Redshift & Classification & $E_{min}$ & $ E_{max}$ & $N_{Spectra}$ & Time Range \\
        &  & & (TeV) & (TeV) & & (MJD) \\
       (1) & (2) & (3) & (4) & (5) & (6) & (7) \\
    \hline
    \multirow{2}{*}{1ES~2344+514} & \multirow{2}{*}{0.044} & \multirow{2}{*}{HBL} &\multirow{2}{*}{0.15} &\multirow{2}{*}{13.00} &\multirow{2}{*}{2} &  (56932 - 56962),(57355 - 57374) \\
    & & & &  &  & \\
    \hline
    \multirow{4}{*}{1ES~1959+650} & \multirow{4}{*}{0.048} &\multirow{4}{*}{HBL} &\multirow{4}{*}{0.30} &\multirow{4}{*}{10.00} &\multirow{4}{*}{10} & (55443 - 55533), (55532 - 56066),(56065 - 56068)\\
    & & & & & & (56067 - 57248),(57247 - 57307),(57306 - 57322), \\
    & & & & & & (57321 - 57423),(57422 - 57530),(57529 - 57550),\\
    & & & & & & (57549 - 57553)\\
    \hline
    \multirow{2}{*}{1ES~1215+303} & \multirow{2}{*}{0.13} &\multirow{2}{*}{HBL} &\multirow{2}{*}{0.11} &\multirow{2}{*}{1.80} &\multirow{2}{*}{3} & \multirow{2}{*}{(56869 - 57040),
(57471 - 57679), (57678 - 57852)} \\
    & & & & & & \\
    \hline
    \multirow{2}{*}{1ES~0229+200} & \multirow{2}{*}{0.14} &\multirow{2}{*}{HBL}&\multirow{2}{*}{0.15} &\multirow{2}{*}{5.10} &\multirow{2}{*}{1} & \multirow{2}{*}{Time averaged} \\
    &  & &  & &  &  \\
    \hline
    \multirow{3}{*}{1ES~1218+304} & \multirow{3}{*}{0.182} &\multirow{3}{*}{HBL} &\multirow{3}{*}{0.15} &\multirow{3}{*}{7.10} &\multirow{3}{*}{8} & (54829 - 54861),(54861 - 54864),(54863 - 54900), \\
    & & & & & & (54899 - 55166),(55165 - 55954),(55954 - 55958), \\
    & & & & & & (55974 - 56313),(56329 - 57518) \\
    \hline
    \multirow{3}{*}{1ES~1011+496} & \multirow{3}{*}{0.212} &\multirow{3}{*}{HBL} &\multirow{3}{*}{0.16} &\multirow{3}{*}{5.10} &\multirow{3}{*}{7} & (54466 - 56690),(56689 - 56693),(56692 - 56694), \\
    & & & & & &(56693 - 56710),(56709 - 56717),(56716 - 56756), \\
    & & & & & &(56755 - 57508) \\
    \hline
    \multirow{2}{*}{1ES~0502+675} & \multirow{2}{*}{0.341} &\multirow{2}{*}{HBL} &\multirow{2}{*}{0.22} &\multirow{2}{*}{1.30} &\multirow{2}{*}{1} & \multirow{2}{*}{(55145 - 55396)} \\
    & & & & &  & \\
    \hline
    \multirow{2}{*}{3C~66A} & \multirow{2}{*}{0.34-0.41} &\multirow{2}{*}{IBL} &\multirow{2}{*}{0.11} &\multirow{2}{*}{0.90} &\multirow{2}{*}{3} & \multirow{2}{*}{(54470 - 54747), (54746 - 54755), (54754 - 57728)} \\
    & & & & & & \\
    \hline
    \multirow{3}{*}{PG~1553+113} & \multirow{3}{*}{0.43-0.58} &\multirow{3}{*}{HBL} &\multirow{3}{*}{0.11} &\multirow{3}{*}{1.26} &\multirow{3}{*}{9} & (55321 - 55350), (55349 - 55671), (55670 - 55874), \\
    & & & & & & (55873 - 56041), (56040 - 56064), (56063 - 56750), \\
    & & & & & & (56934 - 57145), (57144 - 57501), (57500 - 57543)\\
    \hline
    \multirow{2}{*}{PKS~1424+240} & \multirow{2}{*}{0.604} &\multirow{2}{*}{HBL} &\multirow{2}{*}{0.08} &\multirow{2}{*}{2.60} &\multirow{2}{*}{2} & \multirow{2}{*}{(54938 - 55001), (56338 - 57541)} \\
    & & & & & & \\
    \hline
    \end{tabular}{}
    }
    \caption{Data sets considered in this analysis.}
    \label{tab:sources}
\end{table}{}

In this analysis, a subset of the data considered by \cite{elisa} is used.
Only data taken during nominal high voltage level operations were considered.
The data were reanalyzed using a binned-likelihood analysis as discussed in Section \ref{sec:method}.
The data were then binned into nightly-binned time bins and the light curves were obtained.
A Bayesian-block algorithm \cite{scargle} was applied to the data in order to select significant change points in the light curves.
This was performed using the \texttt{astropy} python package \cite{astropy}, using a false alarm probability of 0.01.
All data points, regardless of their significance, were considered when applying the Bayesian-block algorithm.
This reduces the effects of biasing the data from allowing only positive fluctuations of a random signal.
A threshold of $10\sigma$ is applied to each block.
Blocks failing to meet this threshold are excluded from spectral analysis.
This negates the effect of insignificant data inducing a false change point.
If no significant block is observed, the time-averaged data set is used.

\par
\begin{figure}
    \centering
    \includegraphics[scale = 0.3]{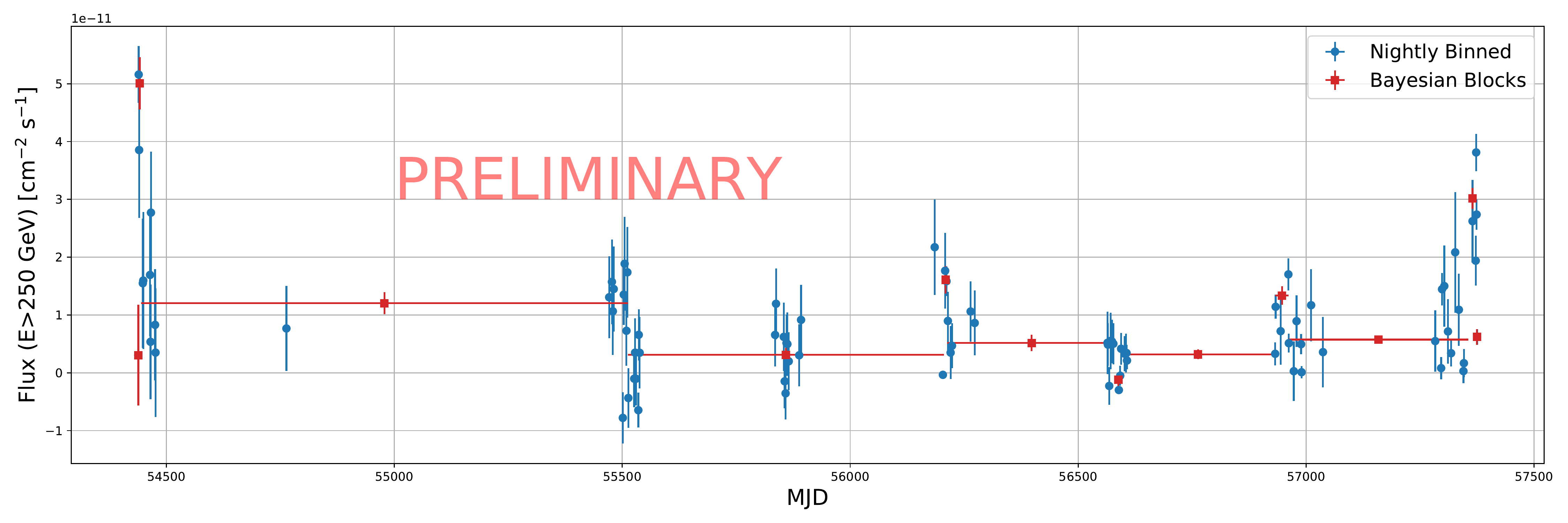}
    \caption{Light curve of 1ES~234+514. The blue circles represent the nightly-binned flux. The red squares represent the average flux obtained for each Bayesian block.}
    \label{fig:1es2344}
\end{figure}

 An example light curve is shown in Figure \ref{fig:1es2344}.
Figure \ref{fig:1es2344} shows the $>250 \mathrm{~GeV}$ light curve for 1ES~2344+514.

In total 12 different blocks are selected. Of the 12 selected, only 3 meet the significance threshold of $10\sigma$.
Observations taken on 54441 MJD were taken under non-optimal weather conditions, hence the block including these data is excluded from the analysis.
The two remaining blocks, and the details of other sources are summarized in Table \ref{tab:sources}.

{}

\par When selecting the optimal model to use when combining profile likelihoods, similar to the method used by \cite{hess_scan}, the $\chi^2$-probabilities are considered.
One chooses the model for which the maximum $\chi^2$-probability is obtained.
In the event of two models having a similar $\chi^2$-probability, the least constraining, i.e. the model with the broadest profile likelihood, is chosen.
This care is taken to prevent biasing the result.

\begin{figure}
    \centering
    \includegraphics[scale = 0.4]{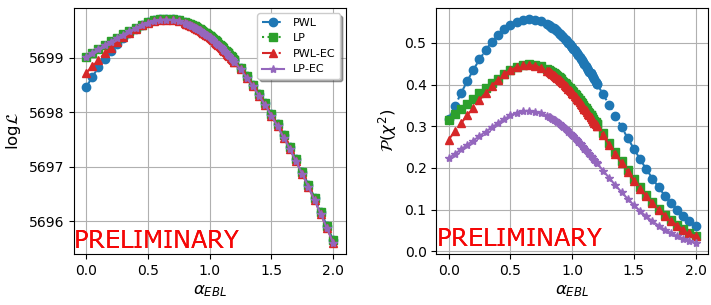}
    \caption{Example of the profile likelihoods (left) and $\chi^2$-probabilities (right) for the different spectral models for 1ES~1011+496 56689-56693 MJD. }
    \label{fig:1es1011_prof_like}
\end{figure}

For example Figure \ref{fig:1es1011_prof_like} shows the profile likelihood and $\chi^2$-probabilities for 1ES~1011+496 56689-56693 MJD.
One can see that the power-law model optimizes the $\chi^2$-probability.
If one were to remove the power-law model, both the log-parabola and power-law with exponential cut-off models obtain a similar $\chi^2$-probability.
If this was the case then the log-parabola model would be chosen as it has a broader profile likelihood.
The effect of model selection tends to only effect the lower bound on the probability distribution as models tend to converge to a power-law model at higher values of $\alpha_{EBL}$, due to constraints on the spectral parameters discussed in Section \ref{sec:method}.
This is evident from the convergence of the profile likelihoods in Figure \ref{fig:1es1011_prof_like} at $\alpha_{EBL} \approx 0.5$.

{}

\section{Results}
\label{sec:results}

The MLE of $\alpha_{EBL}$ for the models tested are given in Table \ref{tab:ebl_results}.
The results show good agreement with the theoretical model ($\alpha_{EBL} = 1$), and, in all cases, provide a significant ($>9\sigma$) improvement to the spectral fits over the no-EBL hypothesis ($\alpha_{EBL} = 0$).

\par
\begin{figure}[!hb]
    \centering
    \includegraphics[scale =0.3]{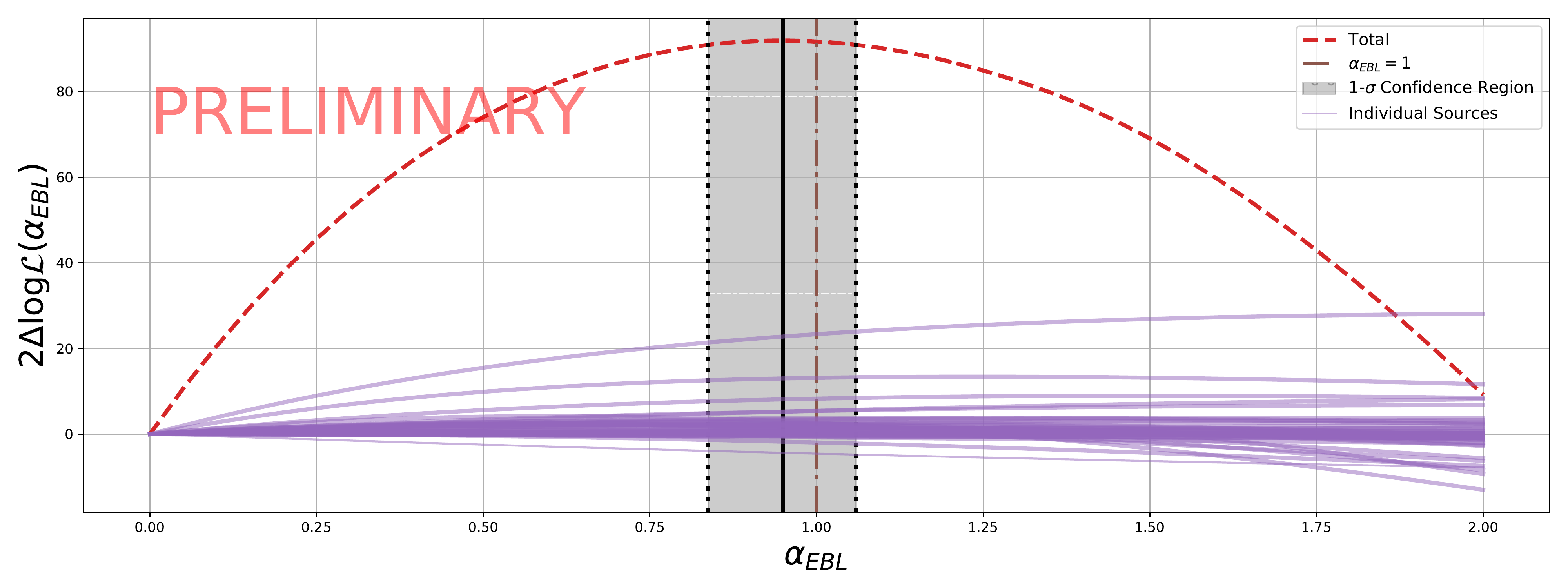}
    \caption{ Combined profile likelihood for a \cite{dominguez11} EBL model. The solid purple lines show the individual source contributions. The dashed red line shows the total combined results. The solid black line shows the MLE of $\alpha_{EBL}$, with shaded gray area denoting the $1\sigma$-confidence interval on the MLE. The brown dashed line shows $\alpha_{EBL} = 1$.}
    \label{fig:combinedResults}
\end{figure}

 Figure \ref{fig:combinedResults} shows the profile likelihood scan as a function of $\alpha_{EBL}$.
The red line represents the summed-profile likelihood for the combined data set.
The gray region is the $1\sigma$ confidence interval on the best fit. The bounds of this region are defined as values of $\alpha_{EBL}$ such that:
\begin{equation}
    2\Delta\log\left(\frac{\mathcal{L}(\alpha_{EBL})}{\mathcal{L}(\hat{\alpha_{EBL}})}\right) = -1.
\end{equation}{}
{}
\par In obtaining the results described in Table \ref{tab:ebl_results}, the lower bound on the redshift estimates for 3C~66A and PG~1553+113 were taken.
To estimate the effect of the uncertainty on the data, and the stability of the fit, the analysis was repeated at the upper bound of the redshift range.
This resulted in a preference for lower values of $\alpha_{EBL}$, and hence only effects the lower-edge of the confidence interval.
To investigate the effect of single spectra dominance, each data set was removed one by one and the analysis was repeated.
The upper and lower bounds of the MLE for each subset are obtained and used to construct a confidence bound.
The uncertainty on the analysis method due to the uncertainty in the redshift of 3C~66A and PG~1553+113 and due to single spectra dominance were combined in quadrature and are reported in Column (2) of Table \ref{tab:ebl_results} with the subscript ``method''.

\begin{table}[tbp]
    \centering
    \resizebox{0.4\textwidth}{!}
    {
    \begin{tabular}{|c|c|c|c|}
        \hline
        Model & $\alpha_{EBL}$ & $\Delta\log\mathcal{L}(\hat{\alpha_{EBL}})$ & $\sigma$ \\
        (1) & (2) & (3) & (4) \\
        \hline
        \multirow{2}{*}{Fi10} & \multirow{2}{*}{$0.98^{(+0.11)_{stat} (+0.17)_{method}}_{(- 0.12)_{stat}(- 0.22)_{method}}$} & \multirow{2}{*}{90.86} & \multirow{2}{*}{9.53} \\
        & & & \\
        \multirow{2}{*}{D11} & \multirow{2}{*}{$0.95^{(+0.11)_{stat}(+0.16)_{method}}_{(- 0.11)_{stat}(- 0.20)_{method}}$} & \multirow{2}{*}{91.88} & \multirow{2}{*}{9.59} \\
        & & & \\
        \multirow{2}{*}{F08} & \multirow{2}{*}{$1.02^{(+ 0.12)_{stat}(+0.17)_{method}}_{(-0.12)_{stat}(- 0.24)_{method}}$} & \multirow{2}{*}{93.02} & \multirow{2}{*}{9.64} \\
        & & & \\
        \multirow{2}{*}{G12} & \multirow{2}{*}{$1.06^{(+ 0.12)_{stat}(+0.18)_{method}}_{(-0.13)_{stat}(- 0.18)_{method}}$} & \multirow{2}{*}{92.59} & \multirow{2}{*}{9.62} \\
        & & & \\
        \hline
    \end{tabular}
    }
    \caption{MLE normalization for various tested EBL models. Column (1) shows the tested EBL model. Column (2) shows the MLE of $\alpha_{EBL}$ for each model and the uncertainty. Column (3) shows the $\Delta\log\mathcal{L}$ at $\hat{\alpha_{EBL}}$ with respect to $\alpha_{EBL} = 0$. Column (4) shows the improvement of the significance with respect to $\alpha_{EBL}=0$. }
    \label{tab:ebl_results}
\end{table}{}

\section{Conclusions}
\label{sec:discus}

\begin{figure}
    \centering
    \includegraphics[scale = 0.325]{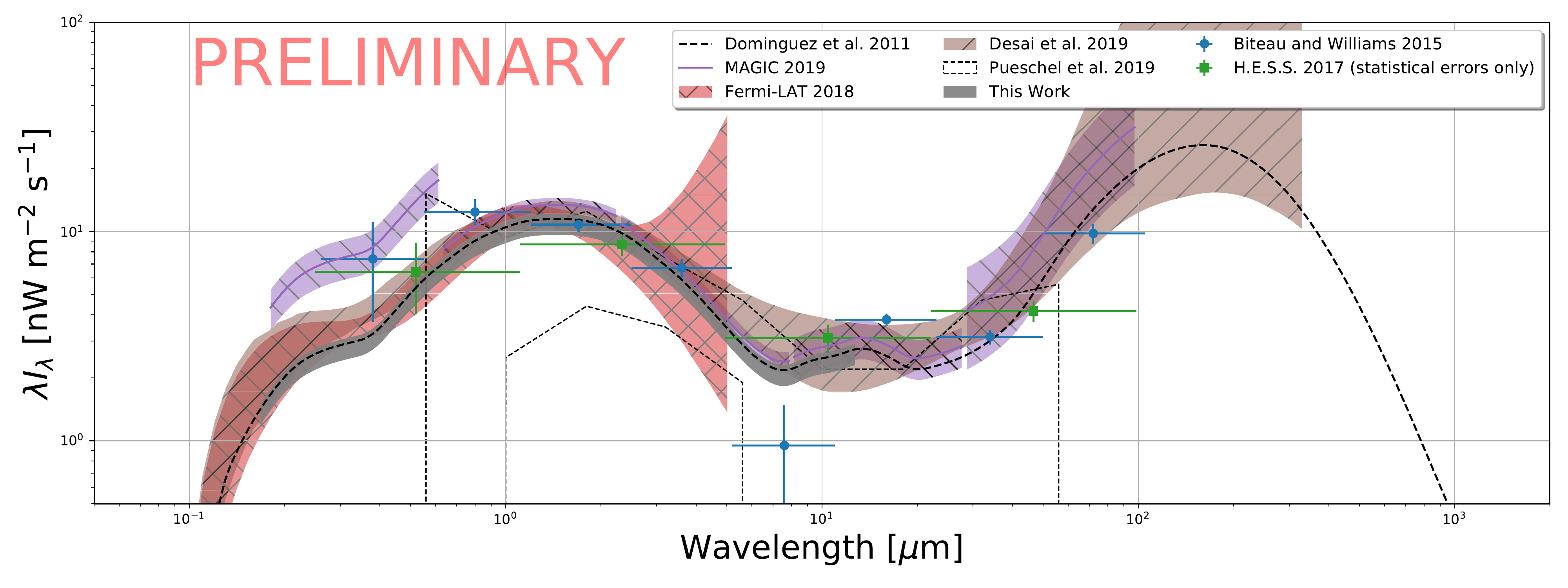}
    \caption{Gamma-ray based measurements of the EBL intensity. Results obtained in the this work for a \cite{dominguez11} EBL model are shown as a gray shaded region. Only statistical errors are shown.
    }
    \label{fig:combinedEBL}
\end{figure}{}

In this work, constraints on the EBL intensity are obtained using VERITAS data.
This analysis utilized the profile and redshift evolution of known EBL models, resulting in model dependant fits to the intensity of the EBL at $z =0$, as reported in Table \ref{tab:ebl_results}.
Figure \ref{fig:combinedEBL} shows results obtained in this study, with other results obtained using HE and VHE instruments.
VERITAS detects the imprint of the EBL on VHE spectra at the $>9\sigma$ level.
The best-fit EBL models are in good agreement with the model-predicted EBL intensity ($\alpha_{EBL} = 1$), leaving little room for unresolved sources.

\par The software developed for this analysis shall allow for future studies of the EBL to be conducted.
Model independent analysis, such as those described by \cite{hess_direct,biteau,fermi}, shall be considered in a future publication.

\section*{Acknowledgement}

This research is supported by grants from the U.S. Department of Energy Office of Science, the U.S. National Science Foundation and the Smithsonian Institution, and by NSERC in Canada. This research used resources provided by the Open Science Grid, which is supported by the National Science Foundation and the U.S. Department of Energy's Office of Science, and resources of the National Energy Research Scientific Computing Center (NERSC), a U.S. Department of Energy Office of Science User Facility operated under Contract No. DE-AC02-05CH11231. We acknowledge the excellent work of the technical support staff at the Fred Lawrence Whipple Observatory and at the collaborating institutions in the construction and operation of the instrument.

\end{document}